# Conflating point of interest (POI) data: A systematic review of matching methods

Kai Sun [1,2], Yingjie Hu* [2], Yue Ma [2], Ryan Zhenqi Zhou [2], Yunqiang Zhu* [1,3]

[1] State Key Laboratory of Resources and Environmental Information System, Institute of Geographic Sciences and Natural Resources Research, Chinese Academy of Sciences, Beijing, China

[2] GeoAI Lab, Department of Geography, University at Buffalo, Buffalo, NY, USA

[3] Jiangsu Center for Collaborative Innovation in Geographical Information Resource Development and Application, Nanjing, China

* Correspondence: yhu42@buffalo.edu; zhuyq@igsnrr.ac.cn

**Abstract**: Point of interest (POI) data provide digital representations of places in the real world, and have been increasingly used to understand human-place interactions, support urban management, and build smart cities. Many POI datasets have been developed, which often have different geographic coverages, attribute focuses, and data quality. From time to time, researchers may need to conflate two or more POI datasets in order to build a better representation of the places in the study areas. While various POI conflation methods have been developed, there lacks a systematic review, and consequently, it is difficult for researchers new to POI conflation to quickly grasp and use these existing methods. This paper fills such a gap. Following the protocol of Preferred Reporting Items for Systematic Reviews and Meta-Analyses (PRISMA), we conduct a systematic review by searching through three bibliographic databases using reproducible syntax to identify related studies. We then focus on a main step of POI conflation, i.e., POI matching, and systematically summarize and categorize the identified methods. Current limitations and future opportunities are discussed afterwards. We hope that this review can provide some guidance for researchers interested in conflating POI datasets for their research.

**Keywords**: Point of interest, POI, POI matching, POI conflation, machine learning, urban studies.

## 1. Introduction

Point of interest (POI) data provide digital representations of places in the real world which may be of interest to some population groups (Psyllidis et al., 2022). Examples of POIs include restaurants, grocery stores, parks, gas stations, fitness centers, schools, hospitals, and government offices. Some POIs are located in natural regions (De Sabbata et al., 2021), while many others are located in urbanized areas and play important roles in supporting various human activities. Without POI data, it would be difficult to represent places and human-place interactions in computational models, especially those models focusing on urbanized areas. Accordingly, POI data have been increasingly used in smart city projects to build digital twins and support urban management decisions (Bilal et al., 2020; Adreani et al., 2022; Ferré-Bigorra et al., 2022).

There exists much research that has utilized POI data to understand urban environments and address urban issues. Leveraging POI type information, researchers conducted research to identify urban functional zones (Gao et al., 2017; K. Liu, Yin, et al., 2020; Ponce-Lopez & Ferreira Jr, 2021; Qian et al., 2021), analyze the landscape of local restaurants (M. Wu et al., 2021; Liang & Andris, 2022), and classify urban land use types (Jiang et al., 2015; R. Wu et al., 2021; Xu et al., 2022). Based on POI locations and human mobility data, researchers conducted studies to infer travel purposes (Y. Liu et al., 2015; Gong et al., 2016; Sari Aslam et al., 2021) and estimate housing values (Fu et al., 2014; Kang et al., 2021). Based on the time when people visit POIs, researchers studied temporal dynamics of cities and regional variability across different urban areas (McKenzie, Janowicz, Gao, & Gong, 2015; McKenzie, Janowicz, Gao, Yang, et al., 2015; Sparks et al., 2020). Based on POI reviews and POI names, researchers conducted studies to derive urban neighborhood representations (Olson et al., 2021), identify urban places to support personal relationships (Bendeck & Andris, 2022), and understand the role of place names in preserving local cultures (Hu & Janowicz, 2018). POI data were also used in many other topics, including understanding urban cognitive places (K. Liu, Qiu, et al., 2020), examining place-related public health issues (Benita et al., 2019; Chang et al., 2022), and measuring urban vibrancy (Tu et al., 2020; Z. Wang et al., 2022). In addition, POI data are utilized to power location-based services (LBS) and location-based social networks (LBSN) (Huang et al., 2021), such as to help users find nearby restaurants or the places visited by friends (McKenzie et al., 2014; Schiller & Voisard, 2004), or to recommend personalized tourist attractions (Santos et al., 2019).

Many companies, organizations, and communities have developed their POI datasets with different geographic coverages and attribute focuses. Commercial companies, such as Yelp, Google, Foursquare, Facebook, and SafeGraph, have all developed their own POI datasets largely to support their own businesses. Many companies require data license purchase for third parties to use their POI data, but some companies have opened their data for academic research without a charge, such as the Yelp Open Dataset[2] and the SafeGraph Core Places data[3]. Non-profit organizations and government agencies may also create and maintain their POI datasets. For example, the US Census Bureau maintains a point-based landmark dataset[4] as part of its TIGER/Line shapefiles, which includes POIs such as airports, cemeteries, parks, mountain peaks/summits, schools, and churches. While published by an authoritative agency, this landmark dataset does not aim to provide a complete coverage of places, and the Census also indicates on their website that they "made no attempt to ensure that all instances of a particular feature were included". The OpenStreetMap (OSM) community has devoted great efforts toward creating and maintaining POI data. The POI data on OSM are generally considered as volunteered geographic information (VGI), and are subject to typical data quality issues (Touya et al., 2017; Yeow et al., 2021). However, a main advantage of the OSM data is that they are free to use, as long as credits are properly attributed. In addition, OSM POI data are available worldwide, although they have different levels of completeness and positional accuracy (Hecht et al., 2013; Zheng & Zheng, 2014).

From time to time, researchers may want to merge two or more POI datasets in order to obtain a better representation of the places in their study areas. This process is generally called *POI conflation* (Low et al., 2021; McKenzie et al., 2014). A main reason for POI conflation is that different POI datasets may have different attribute focuses, place coverages, and data quality. Conflating two or more POI datasets therefore allows us to use these different datasets more effectively by making them complement each other.

We identify three common situations under which POI conflation can be helpful. First, two POI datasets cover places that overlap partially. For example, we may have two POI datasets, each of which covers some but not all restaurants in a similar geographic area (see Figure 1(a)). Under such a situation, POI conflation can help us achieve an increased coverage of places. In addition, the conflation process also combines duplicated POIs (i.e., POI deduplication), and can help reduce potential issues in downstream POI-based analysis (e.g., a computing script may mistakenly count the visits to a grocery store multiple times if there exist duplicated POIs). Second, two POI datasets cover the same places but focus on different attributes. For example, one POI dataset may focus on prices and cuisines of restaurants, while the other POI dataset may focus on reviews (Figure 1(b)). Under such a situation, POI conflation can help increase the comprehensiveness of attributes. Third, one POI dataset has better quality in one or more attributes than the other dataset. For example, the second POI dataset may provide more accurate business-hour information than the first dataset (Figure 1(c)). Therefore, POI conflation can help achieve increased data accuracy. These three situations are not exclusive but can happen in any combination or all together.

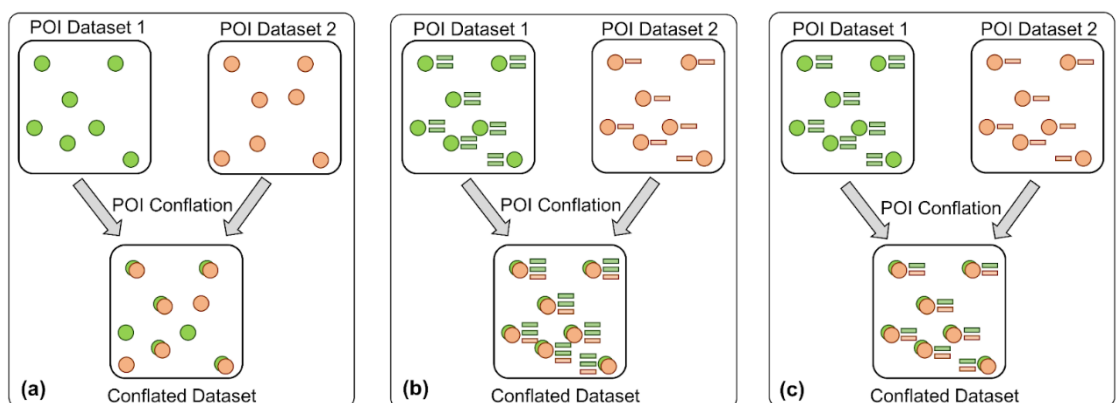

Figure 1. Three common situations under which POI conflation can be helpful: (a) two datasets cover partially overlapping places; (b) two datasets provide complementary POI attributes; (c) one dataset provides more accurate attributes than those of the other dataset.

Many POI conflation methods have been developed in previous research (Lamprianidis et al., 2014; Low et al., 2021; Piech et al., 2020; Yang et al., 2014). However, there lacks a systematic review and it is difficult for researchers new to POI conflation to quickly grasp and use these methods. This review fills such a gap. Following the protocol of Preferred Reporting Items for Systematic Reviews and Meta-Analyses (PRISMA), we conduct a systematic review by searching through three bibliographic databases, i.e., Web of Science, Scopus, and Google Scholar, using reproducible syntax to identify studies related to POI conflation. We then focus on a main step of conflation, i.e., POI matching, and systematically summarize and categorize the identified methods.

It is worth noting that POI conflation can be considered as part of the more general topic of geospatial data conflation. General geospatial data conflation may focus on combining geospatial data in different data models (e.g., vector vs. raster data models) (Chen et al., 2008; Zhang et al., 2011) or having different geometries (e.g., points, lines, and polygons) (Xavier et al. 2016; Lei 2020). While there already exist good review articles on general geospatial data conflation (Hastings 2009; Ruiz et al., 2011; Xavier et al. 2016; Sun et al. 2019; Vilches-Blázquez and Ramos 2021), POI conflation research has its uniqueness and may benefit from its own review. POIs are typically vector data and the spatial footprints of most POI datasets are points. Accordingly, some methodological details for general geospatial data conflation, such as matching raster data to vector data, may be of less interest to a researcher who is particularly interested in POI conflation. In addition, POI datasets often contain attributes not available in other general geographic datasets, such as customer reviews and website URLs. A review paper on general geospatial data conflation usually does not discuss methods for these special attributes. In this context, we believe that a review article specifically focusing on POI conflation can help researchers interested in this topic to quickly gain a more focused view of existing methods.

The remainder of this paper is organized as follows. Section 2 provides background information about POI datasets and POI conflation. Section 3 describes the PRISMA approach that we employ for conducting this review. In Section 4, we review and categorize similarity measures and classification approaches developed for POI matching, and summarize matching results reported in the literature. In Section 5, we identify limitations of current research and potential future directions, and finally, Section 6 concludes this review.

## 2. Background: POI datasets and POI conflation

A POI dataset typically consists of a number of POIs located in a geographic region, and each POI is associated with a set of attributes. More generally, a POI can be represented using Equation (1):

$$POI = < N, T, F, [A, R, \dots] > \quad , \tag{1}$$

where:

- $N$ refers to POI name. A POI can be associated with not only one official place name but also one or multiple alternative names.
- $T$ refers to POI type. A place type is typically assigned to a POI, such as *restaurant*, *grocery store*, and *hospital*. Different datasets usually define their own POI type hierarchies.
- $F$ refers to POI spatial footprint. In most cases, the spatial footprint of a POI is a point (de Graaff et al., 2013), but other geometries such as polygons and lines could also be used.
- $[A, R, \dots]$ refers to other attributes that may be associated with a POI, such as address $A$, reviews $R$, and website URL. These attributes are especially common for POIs in urban areas. We put them in brackets because these attributes have varied availability across different datasets.

It is worth mentioning that the concept of "POI" is also related to *place of interest* (McKenzie & Janowicz, 2018), *area of interest* (Hu et al., 2015; Mai et al., 2018), and *region of interest* (Zeng et al., 2017; Paul et al., 2021). These related concepts largely keep the *interest* part of the concept but explicitly allow the spatial footprints of POIs to be extended to areas or regions. A sense of place (Tuan, 1977) is also sometimes infused into these concepts. In this review, we focus on the commonly used meaning of POI, i.e., *point of interest*.

Similar to the general process of geospatial data conflation, POI conflation also involves two main steps: POI matching and POI merging (Figure 2). The step of POI matching identifies the POIs from different datasets that represent the same places in the real world. The step of POI merging merges the POIs that are matched in the previous step. While Figure 2 illustrates the process using two datasets, the conflation of three or more datasets can be done in an iterative manner.

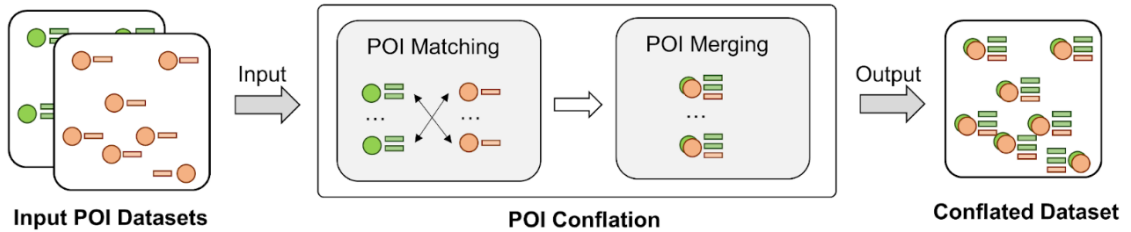

Figure 2. The two steps of POI conflation: POI matching and POI merging.

Most existing POI conflation research focused on the step of POI matching (McKenzie et al., 2014; Piech et al., 2020; Scheffler et al., 2012). A likely reason is that POI matching is a more general step, and methods developed in one study can be applied to other studies too. By contrast, POI merging is a more study-specific step and POI merging rules developed for one study may not be suitable for another study. With this consideration and following the literature, we focus on reviewing POI matching methods in this paper, but also provide a whole picture of the entire process of POI conflation.

## 3. Approach for conducting the review

We conduct a systematic review following the *Preferred Reporting Items for Systematic Reviews and Meta-Analyses* (PRISMA) guidelines (Moher et al., 2009). This approach has been increasingly adopted by researchers in GIScience and urban studies to conduct systematic reviews using clearly specified and reproducible syntax. Examples include systematic reviews on the applications of machine learning approaches to bike-sharing systems (Albuquerque et al., 2021), the role of urban green space in supporting biodiversity (Berthon et al., 2021), the use of unsupervised machine learning and deep learning in urban studies (Grekousis, 2019; J. Wang & Biljecki, 2022), and spatially-explicit GeoAI applications in urban geography (P. Liu & Biljecki, 2022). In the following, we report the methodological details of conducting this review.

### 3.1. Literature searching

We used three bibliography databases to search relevant papers, which are Web of Science[5], Scopus[6], and Google Scholar[7]. These three databases have been widely used by researchers to search for literature, and were also used in previous review articles (Biljecki & Ito, 2021; Fernandes et al., 2020; Vilches-Blázquez & Ramos, 2021). We then used keywords (including key phrases), wildcards, and boolean operators to construct queries to search through these three databases. Specifically, we used the following ten keywords to construct more advanced queries:

> *"Point of Interest", "Points of Interest", POI$, conflating, conflation, matching, match, fusing, fusion,* and *geo\**

The first three keywords limit the topics of the returned literature to those related to POI. Note that *"Point of Interest"* and *"Points of Interest"* are surrounded by double quotes because we want them to be considered as whole phrases. The dollar symbol *$* represents a wildcard for zero or one character, and the keyword *POI$* allows us to find papers that contain the keywords of *POI* and *POIs*. The fourth to ninth keywords, namely *conflating, conflation, matching, match, fusing,* and *fusion*, restrict the methods of the returned literature to conflation, matching, or fusion. Finally, the last keyword *geo\** narrows down the returned literature to those related to *geography* and *geospatial*. The asterisk symbol *\** represents a wildcard for any combination of characters including no character. In addition to the ten keywords, we also tested other keywords, such as *POI\**; however, many irrelevant papers were returned which contain words starting

---

with *poi*, such as *poison* and *poisson*. Thus, we eventually excluded *POI** from the final keyword set. There also exist other keywords that can help identify relevant papers from the broad geospatial data conflation literature, such as *point data*, *places*, or *gazetteer*. While these other keywords could further expand the breadth of this review, they could also dilute its focus on POI data. While we did not include these other keywords in our final searches, interested readers could look into relevant papers, such as Beeri et al. (2004), Hastings, (2008), Li & Goodchild (2011), Xavier et al. (2016), Lei (2020), Acheson et al. (2020), and Lei (2021).

With the keyword set, we connected individual keywords using boolean operators (AND and OR) and parentheses to construct queries which will be used to search through the three bibliography databases. The searches were conducted in the first week of March 2022. The constructed queries were applied to different search fields supported by the corresponding bibliography database, which are "Topic" and "Title, abstract, and keywords" for Web of Science, "Title, abstract, and keywords" and "Anywhere" for Scopus, and "Title" for Google Scholar. We only searched "Title" for Google Scholar because it supports only two search fields (i.e., "Title" and "Anywhere"); and when we tested searching in "Anywhere", it resulted in a huge number of results (e.g., Google Scholar indicated that "about 106,000 results" were found when we applied the first query). Because Google Scholar does not provide any function or Application Programming Interface (API) to download the search results, it is practically impossible to manually go through this huge number of results. Thus, we eventually decided to focus on "Title" only for Google Scholar. Nevertheless, the search results from Web of Science and Scopus already provided a rich set of relevant papers, and the results from Google Scholar further added to this richness.

Table 1 summarizes the used queries, applied data fields, and obtained numbers of returned papers in our searches. We limit the literature language to English and no other limitations were set. Note that because Google Scholar does not support partial matching with wildcards (e.g., the asterisk symbol * will be treated as a place holder of full word) and boolean operators when they are nested, we obtained four sets of results separately using four simpler and non-nested queries (shown in Table 1) and then manually combined the results. A total of 1298 papers were obtained in the final searches based on these three bibliography databases. We assembled the metadata of these papers into a comma-separated values (CSV) file for further analysis, and these metadata include titles, authors, published venues, and years.

Table 1. Queries and the obtained results from the three bibliography databases.

| Database | Queries | Fields | Results |
|---|---|---|---|
| **Web of Science** | ("point of interest" OR "points of interest" OR poi$) AND (conflation OR conflating OR match OR matching OR fusion OR fusing) AND geo* | Title, abstract, and keywords | 179 |
| | | Topic | 199 |
| **Scopus** | ("point of interest" OR "points of interest" OR poi$) AND (conflation OR conflating OR match OR matching OR fusion OR fusing) AND geo* | Title, abstract, and keywords | 266 |
| | ("point of interest" OR "points of interest" OR poi$) AND (conflation OR conflating OR match OR matching OR fusion OR fusing) AND geospatial | Anywhere | 591 |
| **Google Scholar** | "point of interest" AND (conflation OR conflating OR match OR matching OR fusion OR fusing); "points of interest" AND (conflation OR conflating OR match OR matching OR fusion OR fusing); "poi" AND (conflation OR conflating OR match OR matching OR fusion OR fusing); "pois" AND (conflation OR conflating OR match OR matching OR fusion OR fusing) | Title | 63 |

## 3.2. Paper screening

With the initial set of 1298 potentially relevant papers, our next step is to do paper screening in order to narrow down the large number of retrieved papers by filtering out irrelevant ones. As the papers returned from three different databases contain duplicates, we first performed a duplication check by writing a simple Python program to go through the metadata CSV file and detect the papers with the same titles and authors. A total of 433 duplicates were detected and removed in this process, and 865 papers remained. By quickly scanning through the titles of these 865 paper records, we found that there were 13 data records which were not research papers but were conference proceeding volumes (e.g., *14th International Conference on Location Based Services*). These conference proceeding volumes were included because one of their papers contained POI-related keywords, and those papers were already included in the remaining paper records. We therefore removed these 13 conference proceeding volumes, resulting in a remainder of 852 records. Although we specified the language to be English in the search process, there were still 11 non-English papers included in the result. Thus, we removed them. With the remaining 841 papers, we conducted a title screening process with three authors of this review independently reading through these paper titles and labeling out those papers that are clearly irrelevant to POI conflation. Here, we took a conservative approach and only labeled out those papers that are clearly not about POI conflation based on their titles. For example, a paper titled "A Gabor Filter-Based Protocol for Automated Image-Based Building Detection" was labeled as *irrelevant*. Note that such a paper was included in the initial set because its abstract contains "point of interest" and "POI". With the titles independently labeled by three authors, we then used the strategy of majority vote to combine the three labeled results. Papers that were labeled as *irrelevant* by at least two authors were removed, and 552 records were removed through this title screening process. After that, we manually read through the remaining 289 papers primarily focusing on abstracts but also read full paper content if the abstract does not provide sufficient information for determining relevance. In order to identify a final set of core papers for POI conflation and matching, we excluded the following two types of papers:

- *Papers that are related to POIs but do not focus on POI matching or conflation*. These papers often use POIs as one of their geographic data layers to examine social issues, or they may develop software systems for managing or indexing POI data.
- *Papers that are related to matching or conflation but do not focus on matching POI datasets*. These papers are often about matching POIs with non-POI data, such as remote sensing images, or matching POIs with social media users, i.e., POI recommendation.

These two types of papers have their own research merits, but do not focus on matching POI datasets. We therefore removed them (254 papers were removed in this step) and obtained a set of 35 papers.

During the revision of this manuscript in December 2022, we also conducted two sets of additional searches to further expand the literature. First, we conducted four rounds of additional Google Scholar searches by applying the queries to "anywhere" in the article, and manually checked the first 10 pages of the returned search results. Because Google Scholar does not support nested boolean operations, we did four simpler and non-nested queries (thus, four rounds of searches in total). We checked the top 10 pages of each of the four rounds of searches (there are 10 papers on each page and 400 papers in these 40 pages). In the second set of searches, we traced the references of the identified 35 papers to see whether we can identify any additional papers. Through these two sets of searches, we did identify 6 more relevant papers. Among the 6 papers, 4 were published after March 2022 when we did our previous searches, and these 4 papers include 1 journal paper, 2 conference papers, and 1 master's thesis. The 2 other papers that were published before March 2022 include 1 master's thesis and 1 conference paper. With these 6 more papers, we obtained a final set of 41 papers closely focusing on POI conflation.

Figure 3 summarizes our review approach, and a complete list of the 41 papers is provided in the Supplementary. While 41 is not a large number, similar numbers are not uncommon in systematic reviews (Albuquerque et al., 2021; P. Liu & Biljecki, 2022). We also believe that some POI matching and conflation research was done internally by location-based companies, which was probably not shared as research publications. While we do focus on this core set of 41 papers to analyze their methods, our review goes beyond them and we have referred to over 100 papers in this article.

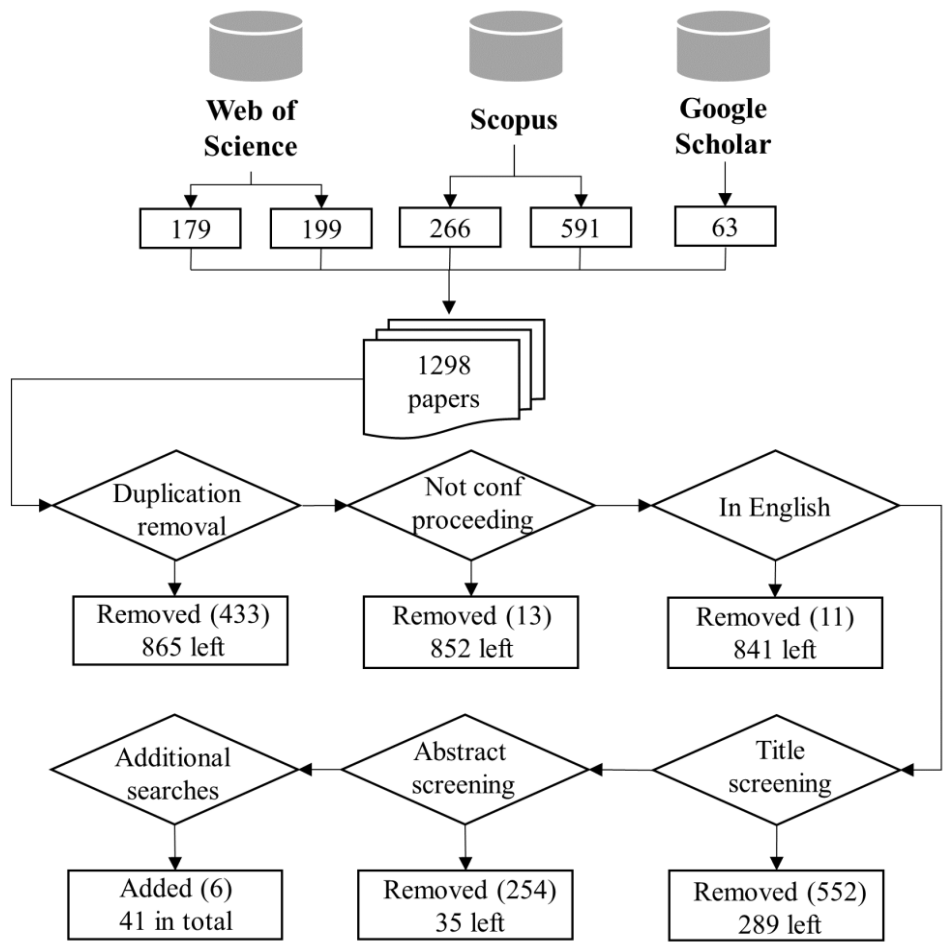

Figure 3. The process of literature searching and screening following the guidelines of the preferred reporting items for systematic reviews and meta-analyses (PRISMA).

## 4. Methods for matching POIs

A common workflow for matching POI datasets is to first measure similarity between any two candidate POIs (or a POI pair) and then use a classification method to classify the POI pair as a match or not match based on the obtained similarity scores. In the following, we review methods for measuring POI similarity in Section 4.1 and classifying POI matches in Section 4.2. After that, we review methods proposed to speed up POI matching for big datasets by generating POI pairs that are more likely to be matches in Section 4.3. In Section 4.4, we discuss evaluation scores achieved in existing POI matching research, and Section 4.5 summarizes this section.

## 4.1. Measuring POI similarity

POI similarity can be measured over POI names, POI types, and spatial footprints. These three main attributes and the geometries associated with spatial footprints have been used in not only POI conflation but also general geospatial data conflation (Hastings 2009; Ruiz et al., 2011; Xavier et al. 2016). In the following, we review similarity measures based on these three main attributes as well as other attributes associated with POIs.

### *4.1.1. Similarity measures over POI names*

High similarity between the names of two POIs is a piece of evidence suggesting that the two POIs may refer to the same place in the real world. Since POI names are generally in the data type of *string*, one approach to measuring POI name similarity is to directly calculate the string similarity of two names. Edit distance has often been used to do so, which measures the similarity of two strings by counting the number of editing operations required to transform one string to another. A smaller number of required operations suggests a higher similarity between two strings. There exist different implementations for edit distance, and each allows a different set of editing operations. The most common one is the Levenshtein distance which only allows the operations of *removal*, *insertion*, and *substitution* (the term "Levenshtein distance" has often been used interchangeably with "edit distance"). Many studies directly used Levenshtein distance for measuring POI name similarity, such as Scheffler et al. (2012), McKenzie et al. (2014), Xia et al. (2014), L. Li et al. (2016), Zhang and Yao (2018), Y. Deng et al. (2019), Barret et al. (2019), Aminy and Lissner (2021), and Zhao et al., (2022). In addition to the Levenshtein distance, other types of edit distance were also used, such as Jaro distance and its variant Jaro-Winkler distance (Barret et al., 2019; Toccu et al., 2019; Piech et al., 2020) and Hamming distance (Barret et al., 2019).

Another approach to measuring POI name similarity is to first apply some text preprocessing steps to POI names and then calculate the similarity based on the preprocessed names. These preprocessing steps include removing non-alphanumeric characters, converting POI names into the same upper or lower cases, filtering out stop words, and tokenizing name strings into individual words. These preprocessing steps generally help remove string differences due to syntactic issues, such as different uses of upper and lower cases or different word orders, and help the similarity measurement focus on the remaining and more meaningful parts of the POI names. Novack et al. (2018) and Low et al. (2021) used *token sort ratio* which tokenizes POI names into individual words, sorts them in alphabetical order to form new strings, and then calculates the Levenshtein distance between the new strings. Similar methods, such as *token set ratio* (based on common tokens in two POI names while ignoring token orders), were also used in Piech et al. (2020). Scheffler et al. (2012) tokenzied POI names into individual words and then used the technique of term frequency and inverse document frequency (TF-IDF) to derive weighted vector representations for POI names; the similarity of POI names was then measured using cosine similarity based on the TF-IDF vector representations. Berjawi (2017) first represented POI names as sequences of tri-grams, and then calculated POI name similarity based on the number of common tri-grams. C. Li et al. (2020) proposed a method that first automatically labels the *semantic roles* of the words in a POI name, and then measures POI name similarity using different weights for words based on their *semantic roles*. The *semantic role* is defined as the function of a word in a POI name. For example, for a POI name "Hilton Garden Inn New York", the semantic role of "Hilton Garden" is *proper name*, that of "Inn" is *common name*, and the role of "New York" is *place name*.

A third approach for measuring POI name similarity is based on the pronunciations of two POI names. Typically, POI names are first converted into sequences of phonetic codes based on their pronunciations using an algorithm such as Double Metaphone. The similarity of the converted phonetic code sequences is then measured using a metric such as the Levenshtein distance. This approach was used by McKenzie et al. (2014). A similar approach was used by L. Li et al. (2016) who leveraged the Microsoft voice packet to transform Chinese POI names into phonetic representations and then measured similarity via the Levenshtein distance.

A fourth approach for measuring POI name similarity is to leverage word embeddings generated by neural network models (S. Wang et al., 2020). Word embeddings are vector representations capturing the meaning of words, and words that share similar meanings are closer to each other in the vector space (Mikolov et al., 2013; Mai et al., 2022). Word embeddings therefore provide a way for measuring semantic similarity of two words even when they have largely different spellings (e.g., "road" and "street"). Similar to word embeddings, character embeddings are vector representations for individual characters, which can help capture the semantics of the strings formed by the characters. Cousseau and Barbosa (2021) trained a deep neural network model that generated both word embeddings and character embeddings to represent POI names, and then concatenated these embeddings for POI matching. Xing et al., (2022) used a pre-trained language model of BERT to encode POI names as vectors. Li et al., (2022) integrated Word2vec and Text-CNN (Kim, 2014) to represent POI names as vector representations.

Table 2 summarizes the four main approaches for measuring POI name similarity and their specific methods. Multiple approaches are often used together in one study (e.g., Albinsson & Sölve (2022) combined multiple name similarity measures from the first, second, and fourth category), and as a result, the same study may show up after different methods in Table 2.

Table 2. Summary of similarity measures over POI names.

| Category | Methods | Studies |
|---|---|---|
| **Similarity based on original POI name strings** | Levenshtein distance | Scheffler et al. (2012); McKenzie et al. (2014); Xia et al. (2014); L. Li et al (2016); Zhang and Yao (2018); Y. Deng et al. (2019); Barret et al. |

| | | (2019); Aminy and Lissner (2021); Zhao et al., (2022) |
|---|---|---|
| | Jaro distance or Jaro-Winkler distance | Barret et al. (2019); Toccu et al., (2019); Piech et al. (2020); Albinsson & Sölve (2022) |
| | Hamming distance | Barret et al. (2019) |
| **Similarity based on preprocessed POI name strings** | Token sort ratio, token set ratio | Novack et al. (2018); Low et al. (2021); Piech et al. (2020) |
| | TF-IDF | Scheffler et al. (2012); Albinsson & Sölve (2022) |
| | Tri-grams | Berjawi (2017) |
| | Semantic roles | C. Li et al. (2020) |
| **Similarity based on POI name pronunciations** | Double Metaphone | McKenzie et al. (2014) |
| | Microsoft voice packet | L. Li et al. (2016) |
| **Similarity based on POI name embeddings** | Word and character embeddings | Cousseau and Barbosa (2021); Xing et al., (2022); Li et al., (2022); Albinsson & Sölve (2022) |

### 4.1.2. Similarity measures over POI types

POI types provide important category information about places. Having the same or similar POI types is another piece of evidence supporting that two POIs may be a match. A main difficulty in measuring POI type similarity lies in the different POI type systems adopted by different datasets. Two POI type systems can have largely different numbers of POI categories (e.g., there were 385 categories in Foursquare and 668 categories in Yelp in the study by McKenzie et al. (2014), different numbers of hierarchies, and different ways to name similar POI types (e.g., two datasets may use two different terms, "*Food*" and "*Restaurant*", to refer to a similar POI type). Accordingly, it is necessary to consider semantic similarity when we compare two POI types.

One approach to measuring the semantic similarity between POI types is to leverage an external lexical resource, such as WordNet. WordNet is a large lexical database that groups words into synsets and links synsets through their semantic and lexical relations (Miller, 1995). As words are linked into a network (with synsets as nodes and semantic/lexical relations as edges), the semantic similarity between two words can be measured using network-based metrics, such as the number of edges between the synsets of the words. WordNet can be used via existing APIs or customized computer programs. McKenzie et al. (2014) used Alignment API 4.0 (David et al., 2011) to measure semantic similarity between POI types using the Path metric. The Path metric is based on the shortest path that connects two synsets in WordNet. Piech et al. (2020) used the library of NLTK to measure semantic similarity based on WordNet, and they used the average of three metrics which are: the Path metric, the Wu–Palmer metric, and the Lin metric. The Wu-Palmer metric is measured based on the depth of two synsets in WordNet and that of their Least Common Subsumer (LCS; or the most specific ancestor node). The Lin metric is measured based on the Information Content (IC; often calculated based on the hyponyms of a synset in WordNet) of the two synsets and that of their LCS. Novack et al. (2018) developed their own program to use WordNet, and they used the average of two metrics, namely the Path metric and the Lin metric. As noted by Novack et al. (2018) and in the literature (Ballatore et al., 2013), the performance of these different metrics largely depends on the specific application and data.

Another approach for measuring the semantic similarity between POI types is to integrate two different POI type systems into one unified system and then measure type similarity based on this unified system. Two POI type systems can be integrated using different methods. Y. Deng et al. (2019) and C. Li et al. (2020) manually integrated two POI type systems from Baidu Maps and Gaode Maps based on the similarity of their top-level POI types. Palumbo et al. (2019) manually linked POI types from eight different sources, including Google, Facebook, and Wikimapia, to OSM tags to form a unified POI type system by considering the activities supported by POI types and the time of these activities. Cai et al. (2021) developed their own 2-level POI type system and manually linked POI types from other datasets to this new POI type. Gong, Gao, and McKenzie (2015) matched the POI types in Foursquare with the POI types in Jiepang (a location-based social media in China) based on POI type names and temporal visiting patterns of people to POI types. Since the unified POI type system is typically in a tree structure, similarity scores between POI types can be calculated using metrics similar to those used in WordNet, such as the number of edges between the corresponding POI type nodes (Barret et al., 2019), similarity score based on common matching ancestors (C. Li et al., 2020; Morana et al., 2014), normalized path distance between two POI type nodes (Y. Deng et al., 2019), and cosine similarity between conception vectors constructed by considering the node depth and descendant density (L. Li et al., 2016).

A third approach for measuring the semantic similarity between POI types is to convert POI types into embeddings via neural network models and then calculate similarity scores based on these POI type embeddings. The rationale is that embeddings can help capture the semantics of POI types and therefore support semantic similarity measurement. One method is to directly use a pre-trained model, such as Sentence-BERT (Silva & Fox, 2021) or fastText (Low et al., 2021), to generate embeddings for POI types. A main advantage of using pre-trained models for generating POI type embeddings is that it does not require additional labeled training data from the current POI conflation project. A main disadvantage is that the generated POI type embeddings may not fit the specific POI datasets in hand. When labeled training data are available, project-specific POI type embeddings can be generated. Cousseau and Barbosa (2021) built their customized encoder model which generated POI type embeddings based on labeled POI pairs. Melo et al. (2022) trained a neural network model based on high-confidence POI matches generated from their previous experiments to learn POI type embeddings. Li et al., (2022) first employed Word2vec to generate encodings for POI types and then used Text-CNN to further refine the vector representations. Once POI type embeddings are generated, their similarity can be measured by cosine similarity (Silva° & Fox, 2021; Low et al., 2021) or dot product (Melo et al., 2022). Table 3 summarizes the three main approaches and their specific methods for POI type similarity measurement.

Table 3. Summary of similarity measures over POI types.

| Category | Methods | Studies |
|---|---|---|
| **Similarity based on an external lexical resource, such as WordNet** | Path metric based on WordNet (via Alignment API 4.0) | McKenzie et al. (2014) |
| | Path metric, Wu–Palmer metric, and Lin metric based on WordNet (via NLTK) | Piech et al. (2020) |
| | Path metric and Lin metric based on WordNet | Novack et al. (2018) |
| **Similarity based on a unified POI type system** | Number of edges between POI type nodes | Barret et al. (2019) |
| | Similarity score based on common matching ancestors | Morana et al. (2014); C. Li et al. (2020) |
| | Normalized path distance between two POI type nodes | Y. Deng et al. (2019) |
| | Cosine similarity between conception vectors constructed with nodes feature | L. Li et al. (2016) |
| **Similarity based on POI type embeddings** | Pre-trained embedding models, such as Sentence-BERT and fastText | Low et al. (2021); Silva° and Fox (2021) |

| | POI type embeddings trained based on labeled POI matches | Cousseau and Barbosa (2021); Melo et al. (2022); Li et al., (2022) |
| --- | --- | --- |

### *4.1.3. Similarity measures over POI spatial footprints*

Spatial proximity of two POIs is another important piece of evidence suggesting that the two POIs may be a match. Given that the spatial footprints of POIs are typically points, one approach for measuring their similarity is through distance calculation between the two points, such as using Euclidean distance (Almeida et al., 2018; Berjawi et al., 2014; N. Wang et al., 2020). Another method is to take into account the curvature of the surface of the Earth and measure geodesic distance (McKenzie et al., 2014; Tré et al., 2013; Zhang & Yao, 2018; Toccu et al., 2019; Zhou et al., 2021; Zhao et al., 2022). As distance measures are often in real numbers, researchers also developed methods to convert real-value spatial distances to similarity measures ranging from 0 to 1. For example, Berjawi (2017) designed an equation based on ellipse function to produce spatial similarity between 0 and 1. Novack et al. (2018) also designed an equation, in which the spatial similarity is calculated as 1 minus the ratio between the Euclidean distance of two points and the maximum possible distance. Y. Deng et al. (2019) used a negative exponential equation to convert the distance between two POIs into the range of [0, 1].

Another approach for measuring the similarity between the locations of two POIs is to leverage embeddings. This approach is primarily used to feed POI distance information to deep neural networks. Cousseau and Barbosa (2021) calculated the geodesic distances between any two POIs and then discretized the distance values of all POI pairs into equal-width bins. They then built a neural network encoder model that generated embeddings for each distance bin, which was then incorporated in a deep learning model for POI matching. Li et al., (2022) used multi-layer perceptron (MLP) to transform the longitude and latitude of POIs into vectors. Table 4 summarizes the two main approaches and their specific methods for similarity measures over spatial footprints.

Table 4. Summary of similarity measures over POI spatial footprints.

| *Category* | *Methods* | *Studies* |
| --- | --- | --- |
| **Spatial distance** | Euclidean distance | Berjawi et al. (2014); N. Wang et al. (2020); Almeida et al. (2018) |
| | Geodesic distance | Tré et al. (2013); McKenzie et al. (2014); Zhang and Yao (2018); Toccu et al., (2019); Zhou et al. (2021); Zhao et al., (2022) |

| | Normalized distance similarity | Berjawi (2017); Novack et al. (2018); Y. Deng et al. (2019) |
|---|---|---|
| **Distance embedding** | Discretized distance embedding | Cousseau and Barbosa (2021) |
| | Spatial coordinate embedding | Li et al., (2022) |

### 4.1.4. Similarity measures over other POI attributes

While almost all POI datasets contain POI name, POI type, and spatial footprints, some datasets contain other attributes, such as addresses, customer reviews, website URLs, and phone numbers. When these and other POI attributes are present, they can be leveraged for POI matching and conflation as well.

*POI address*

POI addresses are commonly available in datasets focusing on business establishments in urbanized areas, such as restaurants, grocery stores, and gas stations. Most POI addresses are in the data type of *string*, and their similarity has been treated in ways similar to POI names in the literature (Lin et al., 2020). One approach is to directly measure the string similarity of two addresses using metrics such as the Levenshtein distance and the Jaro-Winkler distance (Charif et al., 2010; Berjawi, 2017; Comber & Arribas-Bel, 2019; Piech et al., 2020; Psaila & Toccu, 2019; Zhao et al., 2022). Some studies also first standardized addresses into the same address format (e.g., using the format of *door number, street name, city, postcode, country*) before measuring their similarity (J. Liu et al., 2013; Morana et al., 2014; Zhou et al., 2021). A second approach is to segment an address into individual elements, such as street name, city name, and postcode, and then calculate the similarity scores over these individual elements (Y. Deng et al., 2019; C. Li et al., 2020; Low et al., 2021; Melo et al., 2022). A third approach is to leverage embeddings by converting addresses into vector representations using a neural network model before measuring its similarity (Cousseau & Barbosa, 2021; Li et al., 2022).

*Reviews*

Reviews from social media users are available in some POI datasets, such as Yelp and Foursquare. McKenzie et al. (2014) measured review similarity for POI matching. The rationale is that if two POIs refer to the same place in the real world, social media users may discuss similar topics about this POI. To measure the similarity between the reviews of POIs, McKenzie et al. (2014) used an unsupervised topic model, Latent Dirichlet allocation (LDA), to discover topics discussed in user reviews from Foursquare and Yelp, and then leveraged the algorithm of Jensen-Shannon divergence to compute review similarity between two POIs. As reported by the authors, 61.7% of the POIs can be correctly matched purely based on publicly available reviews using their approach (McKenzie et al., 2014).

*Website URL, phone number, and others*

Various other attributes may exist in the POI datasets. For attributes in the data type of *string* (e.g., website URLs and phone numbers), string-based similarity measures, such as the Levenshtein distance, can be applied in ways similar to measuring POI name similarity. For attributes in the data type of *integer* or *float*, numeric value differences, such as absolute difference, can be calculated for measuring their similarity. A number of studies leveraged these additional attributes for POI matching and conflation (Morana et al., 2014; Berjawi, 2017; Almeida et al., 2018; Piech et al., 2020), and they found that additional POI attributes, particularly website URLs and phone numbers, can help effectively identify POI matches when they are present in the POI datasets. Table 5 summarizes the methods for measuring POI similarity over other attributes.

Table 5. Summary of similarity measures over other POI attributes.

| *Attribute* | *Methods* | *Studies* |
|---|---|---|
| **POI address** | Similarity based on string measures, such as Levenshtein distance | Charif et al. (2010); Berjawi (2017); Comber and Arribas-Bel (2019); Psaila and Toccu (2019); Piech et al. (2020); J. Liu et al. (2013); Morana et al. (2014); Zhou et al. (2021) |
| | Similarity based on POI address elements | Y. Deng et al. (2019); C. Li et al., (2020); Low et al., (2021); Melo et al., (2022) |
| | Similarity based on POI address embeddings | Cousseau and Barbosa (2021); Li et al., (2022) |
| **Reviews** | Latent Dirichlet allocation | McKenzie et al. (2014) |
| **Website URL and phone number** | Similarity based on string measures, such as Levenshtein distance | Morana et al. (2014); Berjawi (2017); Almeida et al., (2018); Piech et al. (2020) |

## 4.2. Classifying POI matches

With similarity scores calculated over different attributes, we can then combine these similarity scores to determine whether two POIs are a match. Generally, similarity scores over multiple attributes, instead of a single attribute, need to be considered, because matching errors can easily happen if we consider only one single attribute. For example, considering POI name similarity only can lead to different chain stores under the same name, such as different McDonald's, being classified as a match. Likewise, considering spatial proximity only can lead to different stores located near each other, e.g., different stores located in the same plaza, being classified as a match. There exist two main types of approaches for classifying POI matches: rule-based approaches and machine learning-based approaches. In the following, we review these two types of approaches respectively.

### *4.2.1. Rule-based approaches*

In rule-based approaches, researchers define a number of rules to determine POI matches. One approach is to define rules with crisp matching thresholds for individual attributes, and then combine matching results obtained based on these individual attributes. For example, Scheffler et al. (2012) considered two POIs to be a match if their spatial distance is smaller than 0.01 degree and their POI name dissimilarity (measured by the Levenshtein distance) is less than 10% of the length of the target POI. Ennis et al. (2013) considered two POIs as a match if their POI name similarity is larger than 0.8 (based on a Levenshtein distance measure normalized to [0, 1]), and they have matched POI types, and they are within 20 nearest neighbors based on their spatial distance. Xia et al. (2014) considered two POIs to be a match if they are the nearest neighbor based on spatial distance, and their name and address similarity is higher than 0.6. Similar POI matching rules were also defined in other studies (Lamprianidis et al., 2014; F. Yu et al., 2016). In addition to crisp threshold values, fuzzy logic and fuzzy set theory were also used by researchers to determine POI matches (De Tre & Bronselaer, 2010; Psaila & Toccu, 2019; Toccu et al., 2019).

Another rule-based approach is to first combine different attribute similarity scores into a final score, and then classify a POI pair based on this final score. In this approach, rules are defined to determine the weights of different similarity scores and to decide how the final score should be used for matching. There exist three main types of rules for determining the weights of similarity scores. The first type is that the weights can be determined based on certain assumptions or domain knowledge. For example, Morana et al. (2014), McKenzie et al. (2014), and Berjawi (2017) simply assigned equal weight to each similarity score. Barret et al. (2019) assigned a weight of 0.7 to POI name similarity and a weight of 0.3 to spatial distance similarity, assuming name similarity is more important. A second type of rules is that the weights can be assigned based on relations between the target attributes and other attributes. Novack et al. (2018) designed a method to dynamically calculate the weight for each attribute based on its similarity score over all attributes obtained for the POI pair. Y. Deng et al. (2019) and Zhao et al., (2022) used the Dempster–Shafer evidence theory and analytic hierarchy process as methodological frameworks to calculate weights for attribute similarity scores. The third type of rules is that the weights can be tuned based on a small set of labeled POI pairs (e.g., 200 POI pairs), if such labeled data are available. Note that this approach is different from the machine learning-based approaches that will be discussed in the following subsection, since the tuning process here is about testing different weights in a computational manner, instead of training a machine learning model. For example, N. Wang et al. (2020) used a grid search algorithm to determine the weights for POI name similarity and spatial distance similarity by testing different weights ranging from 0.1 to 0.9. L. Li et al. (2016) developed a method based on information entropy theory to evaluate the entropy of each attribute based on 300 labeled POI pairs, and then calculated the weights of attribute similarity based on the obtained entropy values. Once similarity scores are combined into a final score, POI matches can be determined based on rules, such as the candidate with the highest final score (McKenzie et al., 2014; Morana et al., 2014) or a threshold value based on the final score (L. Li et al., 2016; Melo et al., 2022).

The main advantage of rule-based approaches is that they either do not require any labeled data for training POI matching models or only require a small dataset for weight tuning. The main disadvantage is that it can be difficult to find the most suitable weights and attribute thresholds for these rules. Knowledge about the POI datasets and some trial and error may be needed. Table 6 summarizes the discussed rule-based approaches and their methods.

Table 6. Summary of rule-based classification approaches and methods.

| *Category* | *Methods* | *Studies* |
|---|---|---|
| **Rules defined for individual attribute similarity scores** | Rules with crisp threshold values | Scheffler et al. (2012); Ennis et al. (2013); Xia et al. (2014); Lamprianidis et al. (2014); Yu et al. (2016) |
| | Rules based on fuzzy logic and fuzzy set theory | De Tre and Bronselaer (2010); Psaila and Toccu (2019); Toccu et al., (2019) |
| **Rules defined for combining attribute similarity scores into a final score** | Weights assigned based on assumptions or domain knowledge | Morana et al. (2014); McKenzie et al. (2014); Berjawi (2017); Barret et al. (2019) |
| | Weights assigned based on relations between the target attributes and other attributes | Novack et al. (2018); Y. Deng et al. (2019); Zhao et al., (2022) |
| | Weights tuned based on a small labeled dataset | L. Li et al. (2016); N. Wang et al. (2020) |

### 4.2.2. Machine learning-based approaches

When sufficient training data are available, machine learning models can be trained to combine multiple attribute similarity scores and match POIs. The training data are typically in the form of POI pairs labeled as matching (positive samples) and non-matching (negative samples). A machine learning model takes the similarity scores of different attributes as input features, and learns the weights and classification thresholds through the training process. Trained models can then be used for classifying unseen POI pairs.

A variety of machine learning models have been used in existing studies. For example, McKenzie et al. (2014) trained a binomial probit regression model to combine multiple similarity scores to determine POI matches. Zhou et al. (2021) used a similar approach but trained a logistic regression model. Zhang and Yao (2018) trained a particle swarm optimization (PSO) model for tuning attribute similarity weights by optimizing an objective function for POI matching. Almeida et al. (2018) leveraged the isolation forest model to achieve POI matching through outlier detection, in which the model was trained on matched POI pairs and then used to detect POIs representing different places (i.e., non-matching POI pairs) as outliers. Xing et al., (2022) built a binary classifier based on Light gradient-boosting machine (LightGBM) for POI matching with attribute similarity as input features. Cousseau and Barbosa (2021) developed a deep learning model for POI matching with encoders designed for POI name, type, geographic location, and address respectively. The generated embeddings were concatenated and then passed to a feed-forward network to determine whether two POIs were a match. Li et al., (2022) adopted an enhanced sequential inference model (ESIM) (Chen et al., 2016) to first perform local inference of POI pairs and then combine local inference information to realize the determination of POI pair classification.

Some studies trained multiple machine learning models and compared their performances. Piech et al. (2020) used six supervised learning models in three different types, which are k-nearest neighbor from instance-based models, isolation forest, decision tree, and random forest from decision tree-based model, and perception and deep neural network from neural network models, to learn weights for attribute similarity scores. They tested their trained models for POI matching in five different cities throughout the world, and found that random forest performed the best in four out of the five cities. Low et al. (2021) compared the performance of three rule-based approaches and different machine learning-based approaches designed based on three machine learning models, which are random forest, gradient boosting (based on decision tree), and support vector machine. The authors found that machine learning-based approaches overall achieved better performance than rule-based approaches, and the gradient boosting model outperformed other models. Aminy and Lissner (2021) compared the performances of a decision tree model and an artificial neural network model for POI matching, and found that their performances were overall similar.

The main advantage of machine learning-based approaches is that they can automatically learn weights and thresholds for POI matching and therefore do not require manually defined rules. These learned weights and thresholds usually fit the POI datasets well, and therefore machine learning-based approaches often outperform rule-based approaches when training data are available (Low et al., 2021). A main disadvantage is that they require a fairly large size of labeled training data. Based on the literature, there exist three approaches for obtaining labeled training data: (1) manual labeling (L. Li et al., 2016; Low et al., 2021; McKenzie et al., 2014; Psaila & Toccu, 2019); (2) semi-automatic labeling, e.g., automatically obtaining some matching pairs with high confidence and then checking manually (Cousseau & Barbosa, 2021); and (3) automatic data creation via third-party APIs, such as Factual Crosswalk API (Almeida et al., 2018; Piech et al., 2020). However, even for the automatic approach, some manual effort is needed to verify and ensure that the obtained training data is of good quality. Table 7 summarizes these machine learning models and their related studies.

Table 7. Summary of machine learning-based approaches and methods.

| Category | Methods | Studies |
|---|---|---|
| **Regression models** | Binomial probit regression | McKenzie et al. (2014) |
| | Logistic regression | Zhou et al. (2021) |
| **Optimization model** | Particle swarm optimization | Zhang and Yao (2018) |
| **Instance-based models** | K-nearest neighbor | Piech et al. (2020) |
| | Support vector machine | Low et al. (2021) |
| **Decision tree based models** | Isolation forest | Almeida et al. (2018); Piech et al. (2020) |
| | Decision tree | Piech et al. (2020); Aminy and Lissner (2021) |
| | Random forest | Piech et al. (2020); Low et al. (2021) |
| | Gradient boosting | Low et al. (2021); Xing et al., (2022) |
| **Neural network models** | Perceptron | Piech et al. (2020) |
| | Deep neural network with fully connected layers | Piech et al. (2020) |
| | Deep neural network with fully connected layers and encoders | Cousseau and Barbosa (2021); Li et al., (2022) |

## 4.3. Generating candidate POI pairs

Both similarity measurement and matching classification are done based on candidate POI pairs. If one POI dataset has $n$ POIs and the other POI dataset has $m$ POIs, there exist $n \times m$ possible POI pairs which can be an extremely large number (e.g., when both datasets have about 100,000 POIs). From a computational perspective, the computational complexity of examining all possible POI pairs is $O(n^2)$. Consequently, it can take weeks or even months to complete a POI matching task. While we can certainly use high-performance computing frameworks, such as MapReduce and Apache Spark (Dean & Ghemawat, 2008; J. Yu et al., 2015; Salloum et al., 2016), to speed up the matching process, a more commonly used approach is to reduce the number of candidate POI pairs by limiting them to only those that are located within a certain spatial distance. This process is called *blocking* (Cousseau & Barbosa, 2021; Morana et al., 2014), and is also called *cutoff* or *thresholding* in the geospatial data conflation literature (Hastings, 2009; Li & Goodchild, 2011; Lei, 2020; Lei, 2021). Using this approach, Scheffler et al. (2012) generated candidate POI pairs only for those POIs located within a distance of 0.01 degree (about 1000 meters). Morana et al. (2014) used different distance thresholds to generate candidate POI pairs based on the place type of the target POI, e.g., 50 meters for restaurants and 500 meters for parks. Berjawi (2017) also considered different blocking distances to select candidate POI pairs in terms of different POI types, e.g., 100 meters for restaurants or hotels, and 1000 meters for parks. Piech et al. (2020) used a distance threshold of 300 meters for generating candidate POI pairs, while Low et al. (2021) used a distance threshold of 100 meters. Cousseau and Barbosa (2021) used a Geohash-based approach and only generated candidate POI pairs for those located in the same Geohash keys (about 610 meters). Other distance thresholds were also used by researchers (F. Yu et al., 2016; Zhou et al., 2021). Xing et al., (2022) adopted the k-nearest neighbors algorithm to look for the 10 closest candidate POI pairs based on their spatial locations. Overall, there does not seem to be one standard distance threshold for generating candidate POI pairs, and researchers typically set the distance thresholds based on their own POI datasets.

While the above methods have been specifically used in POI conflation literature, there also exist other methods in the general geospatial data conflation literature for generating candidate pairs. One method is *one-sided nearest-neighbor join*, which generates candidate pairs based on the nearest neighbors of point geographic features (Beeri et al. 2004). Such a method, however, could be limited when many POIs are co-located close to each other (e.g., different stores in the same shopping plaza). Beeri et al., (2004) further proposed three more advanced methods, namely a *mutually-nearest method*, a *probabilistic metho*d, and a *normalized-weights method*. These methods can generate candidate pairs for point geographic features by assigning confidence values to geographic features based on their distances and then selecting those pairs whose confidence values are higher than a defined threshold. In addition to point geographic features, methods were also developed for generating candidate pairs for linear and polygon geographic features, such as *cutoff distance* (Li & Goodchild 2011; Lei 2020) and *overlapping area* (Butenuth et al., 2007; Hastings 2008).

### 4.4. Evaluating POI matching results

Existing studies typically leveraged a subset of the reviewed similarity measures and classification methods to match their POI datasets. Four metrics have been commonly used for evaluating POI matching results, which are *precision*, *recall*, *F-score*, and *accuracy*. Their calculations are shown in Equations (2-5):

$$Precision = \frac{tp}{tp + fp} \qquad , \qquad (2)$$

$$Recall = \frac{tp}{tp + fn} \qquad , \qquad (3)$$

$$F - score = 2 \frac{precision \times recall}{precision + recall} \qquad , \qquad (4)$$

$$Accuracy = \frac{tp + tn}{tp + fp + tn + fn} \qquad , \qquad (5)$$

where $tp$ represents *true positive*, $fp$ represents *false positive*, $tn$ represents *true negative*, and $fn$ represents *false negative*. The evaluation is generally done based on individual POI pairs with ground-truth labels, and $tp$ happens when a POI pair is labeled as a match and the model also considers it to be a match. $fp$ happens when a POI pair is labeled as not a match but the model mistakenly considers it to be a match. $tn$ and $fn$ can be interpreted similarly. Thus, *precision* measures the percentage of correctly matched POI pairs among all the POI pairs that a model considers to be matches. *Recall* measures the percentage of correctly matched POI pairs among all the POI pairs that should be matched (based on the ground-truth labels). *F-score* is the harmonic mean of *precision* and *recall*, and *F-score* will be high if both *precision* and *recall* are high and *F-score* will be low if one of the two is low. *Accuracy* measures the percentage of correctly identified POI pairs (containing both $tp$ and $tn$) among all POI pairs. In addition to the four commonly used metrics, we also see the use of area under the curve (AUC) when the output of a model can be adjusted by different threshold values (Piech et al., 2020), normalized Gini coefficient when the model output is a probability distribution (Cousseau & Barbosa, 2021), and Matthew's Correlation Coefficient for describing the confusion matrix (Albinsson & Sölve, 2022).

Table 8 summarizes the POI matching results of 10 recent studies, which include their used datasets and evaluation metrics. As can be seen, most studies reported POI matches with very high quality. For example, Albinsson & Sölve (2022), Li et al., (2022), C. Li et al. (2020), and Zhou et al. (2021) reported over 0.95 *precision*, *recall*, and *F-score*. Xing et al. (2022), Low et al. (2021), and Melo et al. (2022) all reported over 0.95 *accuracy*. Relatively lower scores were reported by Cousseau and Barbosa (2021), but an *F-score* of over 0.8 is still fairly high. One challenge to interpret these results is that these studies were done based on different evaluation datasets with very different sizes. For example, Cousseau and Barbosa (2021) had the largest data size with over 3 million POI pairs, while some other studies used only several hundred POI pairs. Given these largely different datasets, the evaluation scores from different studies cannot be directly compared. A commonly shared POI matching dataset might help compare results across different studies.

Table 8. Summary of 10 recent studies with their datasets and evaluation scores.

| Study | Dataset* | Precision | Recall | F-score | Accuracy | AUC | Gini |
|--------|----------|-----------|--------|---------|----------|-----|------|
| Zhao et al., (2022) | 387 | 0.984 | 0.969 | - | - | - | - |
| Xing et al., (2022) | - | - | - | - | 0.994 | - | - |
| Albinsson & Sölve (2022) | 5,567 | 0.990 | 0.981 | 0.986 | - | - | - |
| Li et al., (2022) | 439,780 | 0.981 | 0.993 | 0.987 | - | - | - |
| Melo et al. (2022) | 1,942 | 0.973 | 0.860 | 0.913 | 0.953 | - | - |
| Cousseau and Barbosa (2021) | 3,606,880 | 0.837 | 0.712 | 0.809 | - | 0.857 | 0.959 |
| Low et al. (2021) | 8,698 | - | - | - | 0.992 | - | - |
| Zhou et al. (2021) | 300 | 0.962 | 0.968 | 0.965 | - | - | - |
| C. Li et al. (2020) | 1,162 | 0.964 | 0.975 | 0.969 | - | - | - |
| Piech et al. (2020) | 100,000 | - | - | - | - | 0.995 | 0.994 |

\* Numbers indicate dataset sizes represented as the numbers of POI pairs contained.

## 4.5. Summary

In this section, we have systematically reviewed, discussed, and categorized methods for measuring POI similarity and classifying POI matches. We have also discussed the common evaluation metrics and summarized matching results achieved by recent studies. As mentioned previously, this review focuses on the first main step of POI conflation, i.e., POI matching, and the second step, POI merging, can be completed by developing merging rules suitable for a particular study. For example, one can use the attributes of one dataset to update those of the other, or can combine their attributes in a complementary manner. In the next section, we discuss the limitations that we have identified based on this systematic review and some of the future directions that may be pursued to address those limitations.

## 5. Limitations and future directions

We identify six limitations and corresponding directions that may be pursued in the near future.

*First, there is a lack of common test datasets for POI matching evaluation.* All of the reviewed POI matching studies are based on their own POI datasets. This is different from some other research topics on computational methods, such as image classification, in which common large datasets, e.g., ImageNet (J. Deng et al., 2009), were created and many different methods were tested and compared on the same datasets. To some extent, the creation of such common test datasets played key roles in advancing related computational methods, such as deep neural networks for image classification (Krizhevsky et al., 2012). The lack of such common datasets for POI matching is probably due to the high cost of labeling POI pairs. Meanwhile, we can start by sharing smaller datasets, e.g., 200 labeled POI pairs. Since these test datasets will be used for evaluation rather than for training models, they do not have to be very large to begin with.

*Second, a benchmarking platform may need to be built for POI matching evaluation.* When developing a new POI matching method, it is often necessary to compare the new method to previous methods. In reality, it is often time consuming to reimplement previous methods. A benchmarking platform that integrates existing POI matching methods and common test datasets (when available) may help reduce the time that researchers have to spend in reimplementing previous methods. In addition, such a benchmarking platform can support the submission of new POI matching methods developed by researchers, return a performance evaluation result automatically, and even maintain a performance dashboard for matching methods. Similar benchmarking platforms already exist for other computational tasks, such as the General Language Understanding Evaluation (GLUE) benchmark (A. Wang et al., 2019) and the Extensible and Unified Platform for Evaluating Geoparsers (EUPEG) benchmark (J. Wang & Hu, 2019). A similar benchmarking platform may also benefit POI conflation research.

*Third, there is a need for ready-to-use POI matching software tools.* Currently, there does not exist an open-source software tool that researchers can directly use to match their POI datasets. Consequently, researchers who simply want to conflate two POI datasets for their research have to implement matching methods themselves. Although some matching methods can be implemented using existing packages relatively easily (such as using the *FuzzyWuzzy* Python package for Levenshtein distance), a ready-to-use POI matching software tool is likely to facilitate POI conflation for many researchers. One challenge in developing such a software is that different POI datasets may have different attributes. However, as we have shown in this review, most POI datasets contain at least names, types, and spatial footprints, which can be built in a POI matching tool as common attributes. In addition, similarity measures based on different data types, such as strings and numeric values, can be added to such a matching tool, so that researchers can use these measures when additional POI attributes (e.g., phone numbers and reviews) are available.

*Fourth, there is a lack of annotation tools to generate training datasets for POI matching.* Through our systematic review, it has become clear that machine learning approaches generally outperform rule-based approaches, when a fairly large set of training data is available. Creating a labeled training dataset, however, requires considerable manual effort. To obtain training data more efficiently, one possible direction is to develop annotation tools that can assist the work of human annotators. Such a tool may contain functions such as automatically generating POI pairs from a POI database, presenting the POI pairs and their attributes to the human annotator, and saving the annotated POI pairs. Such a tool may also provide multilingual support to help create multilingual POI training datasets. Such multilingual datasets can be especially useful for training machine learning models to match POI datasets in different languages, such as those generated by communities speaking both English and Spanish.

*Fifth, POI similarity measurement may be extended to also include spatial context similarity of POIs.* Current research for measuring POI similarity has been mostly focusing on the inherent attributes of POIs (e.g., POI names, types, and spatial footprints), rather than the spatial contexts of POIs, i.e., how one POI is related to other POIs spatially. It has been shown that POIs in similar types often share similar spatial contexts (Yan et al., 2017; Zhai et al., 2019). For example, restaurants tend to co-locate closely with other restaurants, while hospitals or fire stations are unlikely to co-locate with another hospital or fire station since they need to be distributed more evenly in order to better serve a wider geographic area. By analyzing and quantifying the spatial contexts of POIs, we may be able to add one more dimension of similarity to enhance POI matching. Further research, however, is needed to identify the best approaches for quantifying spatial contexts of POIs, including the spatial distances used for defining contexts and how POIs within the spatial context should be counted toward the similarity measurement.

*Sixth, human-place interactions may also be utilized for enhancing POI similarity measurement.* Another potential direction that goes beyond the typical POI attributes used for matching is to measure POI similarity from a human-place interaction perspective. This type of similarity would require the POI datasets to also have human-place interaction information, such as POI visits in anonymized mobile phone location datasets or POI check-ins in social media data. McKenzie et al. (2015), Janowicz et al. (2019) and Sparks et al. (2020) have shown that different POI types tend to have distinct temporal patterns related to the activities supported at these POIs. For example, restaurants tend to have peak visits during lunch and dinner time, while airports tend to have a more evenly distributed visitation pattern over a day. By analyzing and comparing the human-place interaction patterns of two POIs, we may be able to obtain one additional similarity measurement to improve POI matching. Meanwhile, cautions should be taken in this approach, since the human-place interaction patterns of two POI datasets may be derived from two different population groups (e.g., users of two social media platforms) with different behavior patterns.

The above six directions are by no means an exhaustive list, and many other directions could also be pursued as well. For example, we could further improve semantic similarity measurement between POI types by integrating top-down vocabularies and location patterns derived from bottom-up data-driven approaches (Zhu et al., 2016). Similarly, new machine learning models and location encoding techniques (Mai et al., 2022) could be explored for improving POI matching. We hope that these identified future directions can serve as a starting point to help stimulate new ideas for POI matching and conflation research.

## 6. Conclusions

POIs provide digital representations of places in the real world. POI datasets have been increasingly used to understand human-place interactions, support urban management, and build smart cities. Through POI conflation, we can make effective use of multiple POI datasets and obtain a better representation of the places in our study areas. This paper presented a systematic review of POI conflation with a focus on POI matching methods. Following the PRISMA protocol, we searched through three bibliography databases using reproducible syntax, and a core set of 41 articles were identified out of an initial set of 1298 papers and follow-up searches. We then systematically discussed and categorized the POI similarity measures and matching methods developed so far. Evaluation scores achieved by previous studies were discussed afterwards. Based on this systematic review, we also identified current limitations and potential future directions to address these limitations. We hope that this review could serve as a reference of existing methods for researchers interested in conflating POI datasets to answer research questions.